\documentclass[12pt,3p]{elsarticle}
\usepackage{hyperref}
\usepackage{latexsym}
\usepackage{url}
\usepackage{amssymb}
\usepackage{amsmath}
\usepackage{float}
\usepackage{graphics}
\usepackage{graphicx}
\usepackage{epsfig}
\usepackage{psfrag}
\usepackage{color}
\usepackage{slashed}

\usepackage{amsfonts}

\biboptions{sort&compress}

\begin{document}

\hfill {\tt TIFR/TH/17-44}\\[-0.5cm]

\hfill {\tt CERN-TH-2017-260}\\[0.5cm]

\begin{frontmatter}

\title{The bulk Higgs in the Deformed RS Model}

\author[label1,label2]{F. Mahmoudi\fnref{fn1}}
\fntext[fn1]{Also Institut Universitaire de France, 103 boulevard Saint-Michel, 75005 Paris, France.}
\ead{nazila@cern.ch}
\address[label1]{Univ Lyon, Univ Lyon 1, CNRS/IN2P3, Institut de Physique Nucl\'eaire de Lyon UMR5822,\\ F-69622 Villeurbanne, France}
\address[label2]{Theoretical Physics Department, CERN, CH-1211 Geneva 23, Switzerland}

\author[label4,label5]{N. Manglani}
\ead{namrata.manglani@sakec.ac.in}
\address[label4]{Department of Physics, University of Mumbai,
	Kalina, Mumbai 400098, India}
\address[label5]{	Shah and Anchor Kutchhi Engineering College, Mumbai 400088, India}

\author[label6]{K. Sridhar}
\ead{sridhar@theory.tifr.res.in}
\address[label6]{Department of Theoretical Physics, Tata Institute of Fundamental Research, Homi Bhabha Road, Colaba, Mumbai 400 005, India\vspace*{0.5cm}}

\begin{abstract}
Electroweak precision tests allow for lighter Kaluza-Klein (KK) Higgs modes in the
deformed Randall-Sundrum (RS) model than in models with custodial symmetry. The first KK mode of the Higgs ($h_{1}$) in such a model could have a mass as low as 900 GeV.  In this paper, we study the production of $h_{1}$ and its subsequent decay to a $t \bar t$ pair at the Large Hadron Collider (LHC), in the context of the deformed RS model. We have performed a hadron-level Monte Carlo simulation of the signal and the relevant Standard Model background. We present strategies to effectively suppress the huge SM background and provide a signal that is tractable at the future runs of the LHC.
\end{abstract}


\end{frontmatter}

\section{Introduction}
\noindent One of the most appealing solutions to the large hierarchy between the Planck scale and the electroweak (EW) scale is provided by the Randall-Sundrum (RS) Model \cite{Randall:1999ee}. The RS model is a five-dimensional (5D) 
model with a warped geometry given by the following metric: 
\begin{equation}
ds^2=e^{-2 A(y)}\eta_{\mu\nu}dx^{\mu}dx^{\nu}+ dy^2 \;,
\label{e:metric}
\end{equation}
where, $A(y)$ is called the warp factor. The fifth dimension is compactified on an $S^1/Z^2$ orbifold of 
radius $R$ and, located at the orbifold fixed points, $y=0, \pi R$ are two 3-branes: the UV and the IR brane, 
respectively. In the original RS model (RS1), all the Standard Model fields along with the Higgs are localised 
on the IR brane with only the gravitons being UV-localised. There is essentially only one mass scale to begin
with, viz., the Planck scale $M_p$ but scales associated with the IR-localised fields like the electroweak
vacuum expectation value, are naturally warped down and a stable solution to the Planck-weak hierarchy results.
However, in such a model even other scales that ought to be naturally large, such as the ones that suppress 
proton decay or flavour-changing neutral currents or provide the desirably small neutrino masses, are warped 
down. To avoid such issues and with
the subsequent realisation that only the Higgs need be IR-localised to address hierarchy, a new class of RS models 
in which the Standard Model fields  are allowed to propagate in the bulk were developed \cite{Raychaudhuri:2016,
Gherghetta:2010cj,Davoudiasl:1999tf,Gherghetta:2000qt,Pomarol:1999ad,Grossman:1999ra,Chang:1999nh}. Such bulk
models provide us a framework within which to confront experimental observations more realistically. For instance, 
localising fermions at different points in the bulk provides a tractable approach to Yukawa hierarchy 
\cite{Casagrande:2008hr,Huber:2003tu,Bauer:2009cf,Huber:2000ie,Agashe:2004cp}. 

In fact, even the Higgs field need not be exactly localised on the IR brane: the solution to the gauge-hierarchy
problem requires that the Higgs be only close to it. Localising the Higgs in the bulk close to the IR brane is sufficient to solve the hierarchy problem, so it is not mandatory to fix it on the IR brane \cite{Quiros:2013yaa}. 
With a bulk Higgs the mass bounds on the KK gauge boson ($m_{KK}$) reduces from 12 TeV to 7.2 TeV, i.e. 
by a factor of $\sqrt{3} $ \cite{Cabrer:2011mw}. Even in this light the phenomenological studies of the 
Higgs first KK mode are very few  and have not got their due attention as compared to other SM field KK excitations.

The bulk RS models are severely constrained by the oblique $S$ and $T$ parameters. The constraints from the $S$
parameter are weakened by localising fermions in the bulk but those coming from the $T$ parameter need more serious
consideration. Two different bulk models have been proposed to address this issue:
\begin{itemize}

	\item The first, referred to as the custodial model, invokes a bulk local symmetry ($SU(3)_c \times  
SU(2)_L \times SU(2)_R \times U(1)_y$) which, in a manner reminiscent of the global custodial symmetry of the SM, 
ameliorates the fit to the measured $T$ parameter \cite{Agashe:2003zs, Agashe:2006at}. The bound on the lightest 
$m_{KK}$ comes down to about 3 TeV in such models \cite{Davoudiasl:2009cd,Iyer:2015ywa}. Due to the larger gauge
symmetry of this model, the model has a rich spectrum of new particles. Another issue to contend with in such 
models is the non-universal correction to the $Z \rightarrow b \bar b$ vertex induced by the fact that in
order to get the magnitude of the top quark mass right in bulk models, the $(t,\ b)_L$ doublet cannot be
localised too far away from the IR brane. In custodial models, this is done by embedding the 
$(t,\ b)_L$ doublet in an $SU(2)_L \times SU(2)_R $  bidoublet with a special choice of left- and right-quantum
numbers. The bidoublet contains exotic charge $ \pm 5/3$ fermions.
		
	\item The same problem can be solved in the deformed RS model, without an additional symmetry. In this model, 
we assume a bulk Higgs i.e. a Higgs not on the IR brane but close to it and introduce an additional scalar field. 
Due to this extra field, warping of the fifth dimension is strongly modified near the IR brane, while it behaves 
as pure AdS near the UV brane. This is done using soft wall metrics and a naked singularity generated beyond the 
IR brane by this scalar field. Proximity of the IR brane to the singularity determines the strength of the 
modification. The deformation of the metric tends to localise the gauge KK modes closer to the IR brane than in the 
normal RS model and with the Higgs zero mode localised further out in the bulk its overlap with the gauge 
KK modes is reduced. This relaxes the electroweak contraints considerably \cite{Cabrer:2011fb,Iyer:2015ywa}.
	
In addition the $Z\rightarrow b \bar{b}$ partial width and flavour observables also provide stringent constraints 
on the gauge KK mass. However even after taking these into account, lower bounds on $m_{KK} \sim \mathcal{O}$($1 -2$ 
TeV) are obtained \cite{Cabrer:2011qb} in a reasonably significant part of the parameter space of this model, 
making it interesting from the LHC perspective.

\end{itemize}

Given that the deformed RS model is a viable alternative to the actively investigated custodial RS model, it 
is worthwhile to also subject the deformed RS model to a more critical scrutiny, specially from the point of 
view of collider searches. A couple of studies of collider signals in the deformed model have been published 
\cite{Iyer:2016yjb,deBlas:2012qf}, but, other collider signals in the deformed model are crying out for attention.

In this paper, we study the production of the first KK mode of Higgs within the framework of deformed RS model. 
A similiar study for the same process within the custodial RS model was published by us earlier \cite{Mahmoudi:2016aib}. However, the significantly lower mass range available for the first KK mode of the Higgs in the deformed RS model and the much smaller production cross sections as compared to custodial RS model makes the collider analysis 
more challenging. Not only do the lower cross sections pose a challenge but at the lower mass end the Standard Model 
backgrounds also turn out to be a very serious problem. It is to address these challenges that we have to alter the 
analysis from the previously studied custodial case \cite{Mahmoudi:2016aib}.  

The paper is structured as follows: In Section 2 we provide a brief introduction to the deformed RS model along with a brief description of the constraints. In Section 3 we give a detailed explanation regarding the signal and background simulations and the strategies used to suppress the background effectively. In Section 4 we summarise the results.

\section{Bulk Higgs in Deformed RS Model}
The action for a bulk Higgs and other scalar fields ($\phi$) in a 5D theory is given by \cite{Quiros:2013yaa}:

\begin{equation}
S_{5}= \int d^{4}x dy \sqrt{-g}\Bigl[-|D_{M}H|^{2}-\frac{1}{2}|D_{M}\phi|^{2}-V(H,\phi)-\Sigma_{\alpha} (-1)^{\alpha}2 \lambda^{\alpha}(H,\phi)\delta(y-y_{\alpha})\Bigr]\;,
\label{e:action}
\end{equation}
where $\lambda^{\alpha} (\alpha = 0, 1)$ are the brane potentials for the UV and the IR branes respectively, which are of the form $\lambda^{0}(\phi_{0},H)=M_{0}|H|^{2}$ and $-\lambda^{1}(\phi_{1},H)=-M_{1}|H|^{2}+\gamma |H|^{4}$. Here $\phi_{\alpha}$ is the vacuum expectation value of the field $\phi$ at the two boundaries of the fifth dimension $y=y_{\alpha}$.

The $ V(H,\phi) $ is the 5D  scalar potential having a quadratic mass term with the coefficient $M(\phi)$ and $H$ is the 5D Higgs field having the notation:
\[
H(x^{\mu},y)=\frac{1}{\sqrt{2}}
\begin{bmatrix}
0 \\
h(y)+\xi(x^{\mu},y)
\end{bmatrix}\;,
\]
where $h(y)$ is the Higgs background and $\xi(x^{\mu},y)$ can be expanded as a series of the Higgs KK modes. For a small Higgs mass, we can assume that the vacuum expectation value (vev) is almost entirely carried by the zero mode ($h_{0}$), hence the zero mode profile is the same as the vev profile.

The differential equations for the profiles of $h(y)$ and $\xi(y)$ are obtained by varying the  5D action of the scalar fields given in Eq. (\ref{e:action})

\begin{equation}
h''(y)-4A'(y)h'(y)-\frac{\partial V}{\partial h}=0\;,
\label{e:h0}
\end{equation}
with the boundary conditions
\begin{equation}
\frac{h'(y_{\alpha})}{h(y_{\alpha})}=\frac{\partial \lambda^{\alpha}(h)}{\partial h}|y=y_{\alpha}\;.
\end{equation}
Similarly, for $\xi(y)$ we have
\begin{equation}
\xi''(y)-4A'(y)\xi'(y)-\frac{\partial^{2} V}{\partial h^{2}}\xi(y)+m_{n}^{2}~e^{2A}\xi(y)=0\;,
\label{e:kk}
\end{equation}
with the boundary conditions

\begin{equation}
\frac{\xi'(y_{\alpha})}{\xi(y_{\alpha})}=\frac{\partial^{2} \lambda^{\alpha}(h)}{\partial h^{2}}|y=y_{\alpha}\;.
\end{equation}
After simplifying the above differential equations, we can obtain the solutions for the profiles of $h_{0}$ and $h_{1}$. \\

The profile equations for the $h_{0}$ and fermion zero modes ($t^{L,R}_{0}$) as given in Refs. \cite{deBlas:2012qf,Cabrer:2011qb} are
\begin{eqnarray}
f^{h}_{0} &=& N^{h}_{0} e^{a k y- A(y)}\;, \nonumber \\
f^{t_{L},_{R}}_{0}&=&N^{t_{L},_{R}}_{0} e^{(0.5 \mp c )A(y)}\;.
\end{eqnarray}

Using these profile equations we fix the value of the fermion mass parameter ($c$) by fitting the top quark mass \cite{Iyer:2015ywa}. We fit the 5D Yukawa ($y_{5}$) to the SM Yukawa ($y_{4}$) using these profiles ($y_{4}= y_{5} \int^{y_{1}}_{0} f^{h}_{0}  f^{t_{L}}_{0}  f^{t_{R}}_{0} dy$) by multiplying the 5D Yukawa with the profile overlap integral for the profiles of the zero-mode Higgs to the zero-mode left handed top quark and the zero-mode right handed top quark. The coupling modifier ($y_{100}/y_{4}$) for the $h_{1}$ coupling to the zero-mode top quarks is given as the ratio of the profile overlap for KK Higgs first mode with the top quarks to the profile overlap of KK Higgs zero mode with the top quarks ($y_{100}= y_{4} \times \frac{\int^{y_{1}}_{0} f^{h}_{1}  f^{t_{L}}_{0}  f^{t_{R}}_{0} dy}{\int^{y_{1}}_{0} f^{h}_{0}  f^{t_{L}}_{0}  f^{t_{R}}_{0} dy}$).

The main ingredient of the deformed RS model \cite{Cabrer:2011fb,Cabrer:2010si} is the modified metric given in Eq. (\ref{e:metric}), where the warp factor $A(y)=ky$ for the custodial RS model and for the deformed RS model is
\begin{equation}
A(y)= k y - \frac{1}{\nu^2}Log(1-\frac{y}{y_{s}})\;.
\label{e:Ay}
\end{equation}
Here $y_{s}$ denotes the position of the singularity which is at a distance of $\Delta$ beyond the IR brane ($y_{1}$) along the fifth dimension such that $y_{s}=y_{1}+\Delta $. The parameter $\nu$ defines the extent of deformity, $\nu \rightarrow \infty$  being the limit in which this model is like the normal RS model without deformation. The parameter $\Delta$ is the measure of proximity of the IR brane to the singularity. Thus we have $\nu$ and $\Delta$ as free parameters of the model. The $y_{1}$ value is fixed using the constraint $A(y_{1})=36$, which is required to solve the hierarchy problem.  To keep the perturbativity in the 5D theory under control we keep $M_{5} L_{1} \geq 1 $, where $M_{5}$ is the 5D Planck scale ($M_{p}^{2}=M_{5}^{3} \int e^{-2A(y)}dy$) and $L_{1}$ is the curvature radius at the IR brane. Since $kL_{1}=\frac{\nu^{2}k\Delta}{\sqrt{1-2\nu^{2}/5+2\nu^{2}k\Delta +\nu^{4}(k\Delta)^{2}}}$, if we choose $kL_{1} < 1$ we get a parameter set for deformed model that departs from AdS. A smaller value of $kL_{1}$ implies larger deformation. If we select $kL_{1} \geq 0.2$ the hierarchy between $M_{5}$ and k can be restricted from growing too large. The fine tuning parameter $\delta \equiv |f(y_{1})| \lesssim \mathcal{O}(1)$ implies that the Higgs solution is free of fine-tuning, where $f(y)$ is defined in the Eq. (6.5) in Ref. \cite{Cabrer:2011fb}. For this model the coefficient of the quadratic mass term in the scalar potential is $M{(\phi)}=k^{2}\Bigl[a(a-4)-4ae^{\nu \phi / \sqrt{6}}\Bigr]$, where $a$ is the Higgs bulk mass parameter.

\begin{figure}[t!]
	\begin{center}
		\includegraphics[origin=rb,width=3.5in]{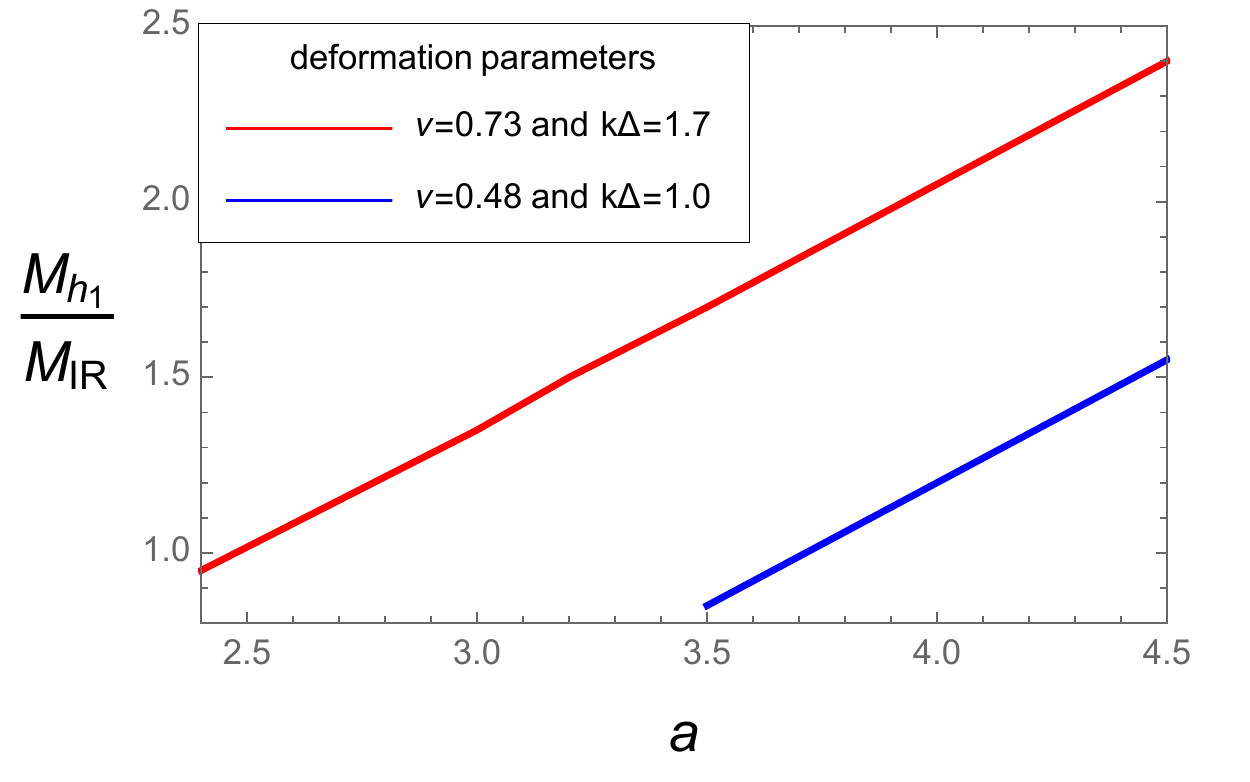}
	\end{center}
	\caption{\it Variation of the KK Higgs mass for its first mode ($M_{h_{1}}$) in terms of the IR scale $(M_{IR})$ with respect to $a$ parameter for two sets of deformed model parameters. } \protect\label{a_mkk}
\end{figure}

The solution to the gauge hierarchy problem demands that the Higgs field zero mode should be localised on/close to the IR brane. This implies that $a \geq a_{0}$, for the RS model without deformation $a_{0}=2$. In case of the deformed RS model, we choose $a_{0}=2A(y_{1})/ky_{1}$ as given in Ref. \cite{Cabrer:2011qb}. The mass of the KK Higgs mode $m_{n}$ depends on the IR scale ($M_{IR}= k~ e^{-A(y_{1})}$) and $a$. The variation of the mass of the KK Higgs first mode ($M_{h_{1}}$) in terms of $M_{IR} $ with respect to the parameter $a$ for two sets of deformation parameters can be seen in Figure \ref{a_mkk}. Thus we show that the lower value of the $a$ parameter for a given set of deformation parameters can be more interesting for the deformed RS model from the point of view of LHC phenomenology. The parameter space that we consider in the following is  $\nu=0.48$ and k$\Delta=1$ and $a=3.2$ \cite{Cabrer:2011fb}.\\

\begin{figure}[!t]
	\begin{center}
		\begin{tabular}{cc}	
			\includegraphics[width=8 cm, height= 5cm]{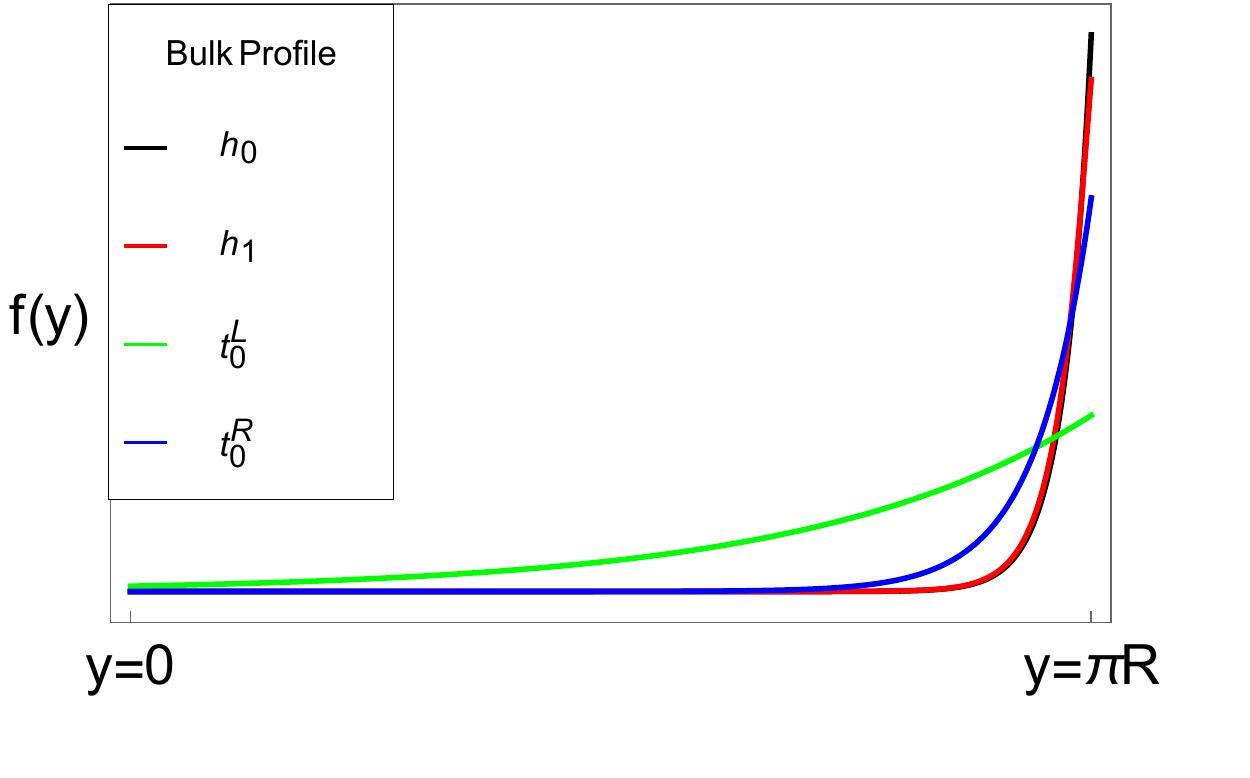}&	
			\includegraphics[width=8 cm, height= 5cm]{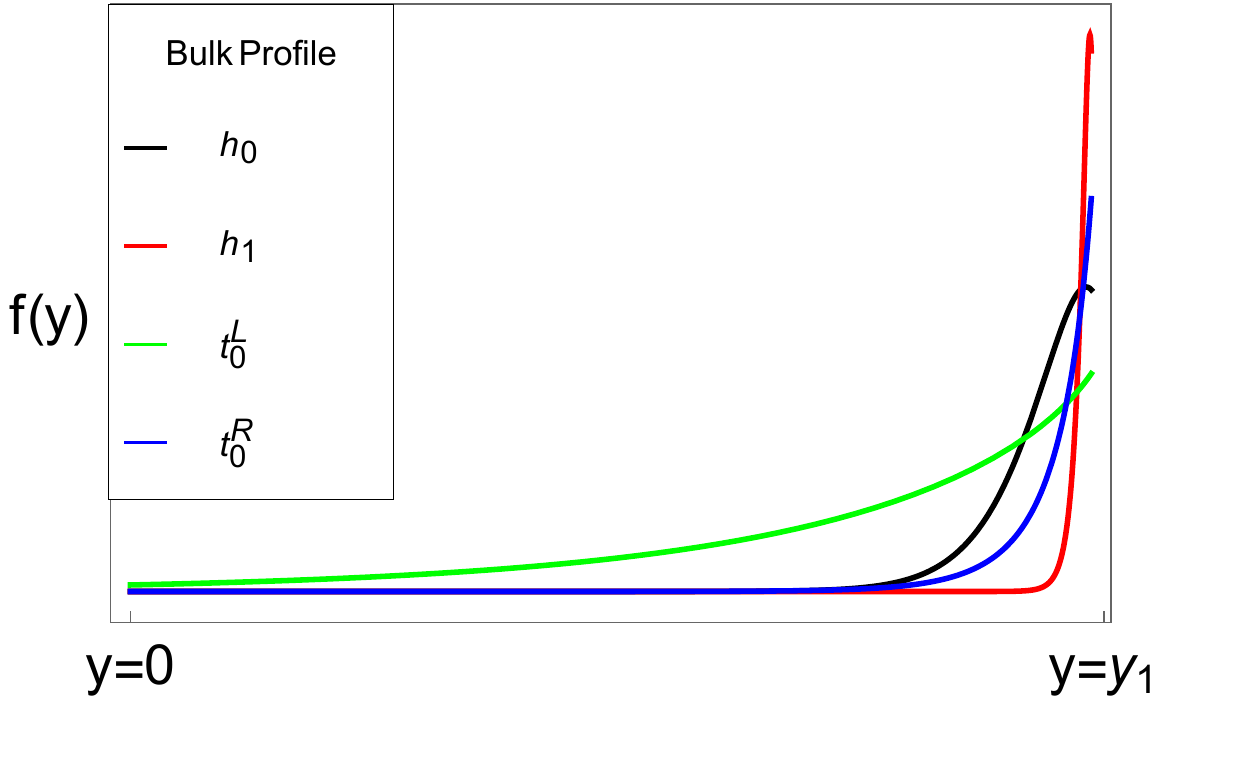}\\
		\end{tabular}
	\end{center}
	\caption{\it Profiles for the first KK mode of Higgs (red), zero-mode of Higgs (black), zero-mode of $t^{R}$ (blue) and zero-mode of $t^{L}$ (green) for the custodial RS model in the left and the deformed RS model in the right.}\label{profile}
	
\end{figure}

In Figure \ref{profile}, we have the plotted profiles  $f^{t_{L}}_{0}$, $ f^{t_{R}}_{0}$, $f^{h}_{1}$ and $f^{h}_{0}$ for the custodial and deformed RS model. We observe that the $f^{h}_{0}$ profile is less IR localised whereas the $f^{h}_{1}$ is more IR localised in the deformed model. This reduces the profile overlap of $h_{1}$ with the $t_{L}$ and $t_{R}$ in the deformed model, resulting into smaller couplings which makes the effective cross section in the deformed model smaller as compared to the custodial case for the same $M_{h_{1}}$.  Hence a separate search strategy is required for the deformed case. Moreover, the softer couplings and the resulting tiny cross sections imply that the existing constrains from direct searches at the LHC have little impact on this model.

\section{Signal and background simulation}

The signal is characterised by a bulk Higgs ($h_{1}$), which is the Kaluza Klein first mode of the Higgs boson decaying to a pair of boosted top quarks in the context of deformed RS model. For this model the coupling of $h_{1}$ to the weak bosons vanishes at leading order due to the orthogonality condition for profiles of $h_{1}$ and $h_{0}$. The production cross section for the $h_{1}$ in association with a top quark pair is very small. Hence, gluon-gluon fusion is the main process for the production of $h_{1}$. The coupling of $h_{1}$ to the top quark pair is nearly equal to the SM top Yukawa coupling. We probe signals with $M_{h_{1}}>$ 800 GeV. At this mass the top decay channel is open and dominant. Thus the signal topology that we are interested to study is as follows:
\begin{equation}
p~p ~(g~g)~\rightarrow~ h_{1}~\rightarrow~t \bar{t}
\end{equation}

The model files for $h_{1}$ were generated using {\tt{FEYNRULES}} \cite{Alloul:2013bka} taking into account the effective coupling of $h_{1}$ to a pair of gluons via a top quark loop. The parton-level amplitudes for the signal were generated using {\tt{MADGRAPH}} \cite{Alwall:2014hca} at 14 TeV centre of mass energy using parton distribution function {\tt{NNLO1}} \cite{Ball:2012cx}, and showering was done in {\tt{PYTHIA 8}} \cite{Sjostrand:2007gs}. The most dominant backgrounds for our signal are $t\bar t$ and QCD. Events for the $t\bar t$ background and the QCD background have been generated directly in {\tt{PYTHIA 8}}. To generate the background events with larger statistics, we choose phase space  cuts specified by $\hat{p_{T}} > 300$ GeV and $\hat{m}\in(M_{h_{1}}-300~{\rm GeV}, M_{h_{1}}+300~{\rm GeV})$ for the mass range of 900 GeV to 1000 GeV, while $\hat{p_{T}} > 400$ GeV and $\hat{m}\in(M_{h_{1}}-300~{\rm GeV}, M_{h_{1}}+300~{\rm GeV})$  are chosen for the mass range of 1100 GeV to 1300 GeV, where the hat represents the outgoing parton system. 

The signal is characterised by a pair of top quarks which come from the decay of massive $h_{1}$ and they tend to be boosted, with their transverse momentum in the range of 200 GeV to 500 GeV. Such a boost will ensure that the top decay products will lie in a single hemisphere. So, we have reconstructed fat jets from final state partons employing  {\tt{FASTJET}} \cite{Cacciari:2006sm,Cacciari:2011ma} and using the Cambridge Aachen (C-A) algorithm \cite{Dokshitzer:1997in,Bentvelsen:1998ug}  for clustering by setting the jet radius parameter $R = 1.5$. We accept only those jets which satisfy $|\eta|<2.7$ and $p_T>300$ GeV.\\

We are interested in hadronic decays of the top quark, therefore we select events which have no leptons that satisfy $p_{T} >25$ GeV and $|\eta|<2.5$. Once we have a hadronic decay of both the top quarks, we need the event to be characterised with two fat jets and each of them should satisfy $|\eta|<2.7$ and $p_T>300$ GeV. These two fat jets are then considered as an input for the {\tt{HEPTopTagger}} \cite{Plehn:2011tg, Kasieczka:2015jma}. This is the most effective top tagger in the momentum range of our interest. Once both the leading and subleading jets pass the {\tt{HEPTopTagger}}, the event is selected for further analysis.

\begin{figure}[!t]
	\begin{center}
		\begin{tabular}{cc}	
			\includegraphics[width=8 cm, height= 5cm]{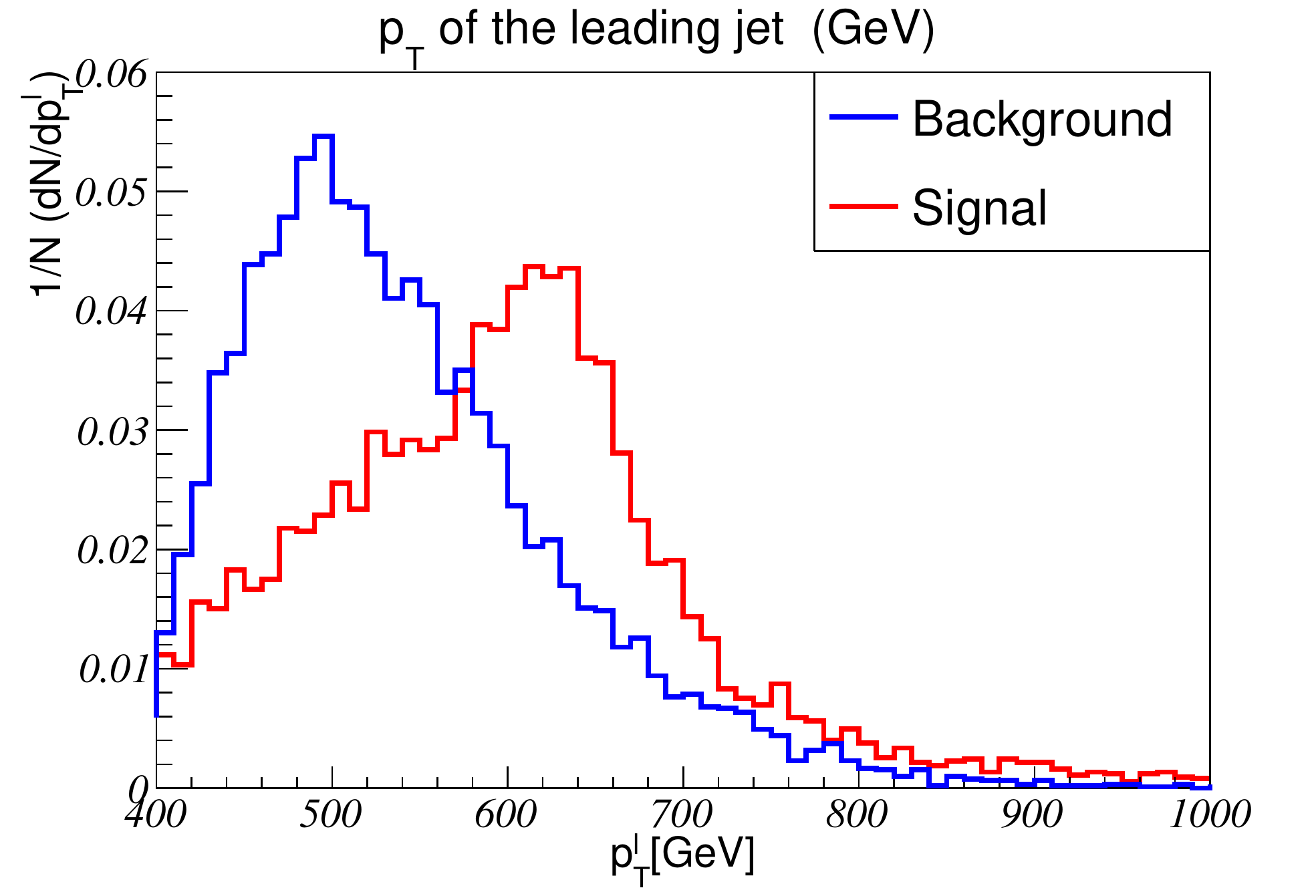}&	
			\includegraphics[width=8 cm, height= 5cm]{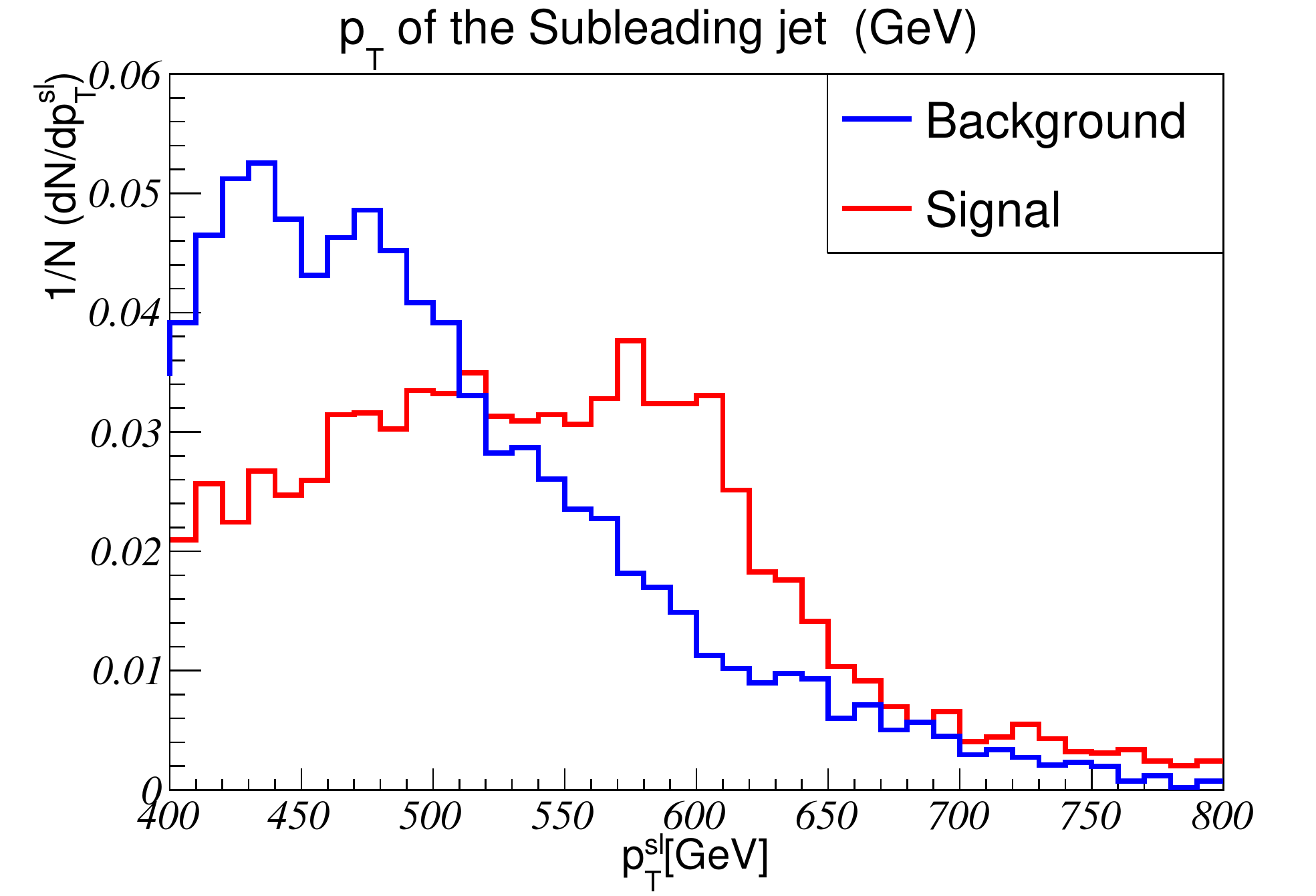}\\
		\end{tabular}
	\end{center}
	\caption{\it Normalised distribution of the $p_{T}$ for leading jet  (left) and subleading jet (right).
		The red distribution represents the signal and the blue one represents the $t\bar{t}$ background for $M_{h_{1}}=1300~\rm{GeV}$.}\protect\label{pt_top_sb}
	
\end{figure}

The use of {\tt{HEPTopTagger}} helps in reducing the QCD background whereas a mass dependent $p_{T}$ cut on the leading jet and the subleading jet helps in curbing both the QCD and SM $t \bar{t}$ backgrounds. The $p_{T}$ cuts for the leading and subleading jets from $M_{h_{1}}=1300$ GeV can be explained by plots shown in Figure \ref{pt_top_sb}. The SM $t\bar{t}$ background largely peaks at comparatively lower transverse momentum. It is thus brought under control using a $p_{T}$ cut on the leading jet ($p_{T}^{l}~>~580$ GeV) and the subleading jet ($p_{T}^{sl}~>~540$ GeV). The QCD background is huge, specially for the invariant mass range of our interest it is very difficult to control it with a $p_{T}$ cut alone. We bring it down by using $b$-tagging inside the fat jet tagged as a top-jet using {\tt{HEPTopTagger}}.

A fat jet which is tagged as a top-jet has three main subjets, two of them reconstruct the $W$ boson mass and the remaining one is a $b$-like subjet. We calculate the $\Delta R$ between this $b$-like subjet and the $b$-quark for both the leading and subleading top-tagged fat jets. Events that give a $\Delta R <0.1$ (tight $b$-tag) for both the leading and subleading top-tagged jets, are referred as double $b$-tag events. Selecting such events after the use of {\tt{HEPTopTagger}} helps to tame the QCD background. We have also taken into account the mistagging probabilities of $c$-quarks (20$\%$) and light quarks(1$\%$) for a $b$-tagging efficiency of 0.7 \cite{CMS:2016kkf}.

We present two sets of cuts, one suited for the lower mass of the $h_{1}$ and the other one for the higher mass of the $h_{1}$,  as shown in Table \ref{table1}.
\begin{table}[htb]
	
	\begin{center}
		\begin{tabular}{|c|c|c|c|c|} \hline
			$M_{h_{1}}$ (GeV)&Cuts& Signal (fb)& $t\bar t$ (fb) & QCD (fb) \\ \hline
			
			900& 2 fat jets with $p_{Tmin}=300$ GeV&  101.22	&4730.86	&6338534.63  \\
			&2 top-tagged jets& 10.72	&553.77	&3641.83\\
			&$p_{T}^{l}>380$ GeV and $p_{T}^{sl}>360$ GeV&	7.28	&264.62	&2446.48\\
			& $b$-tagging for both the jets &3.09	&118.12	&0\\
			&$900~{\rm{GeV}}<m_{tt}<1000~{\rm{GeV}} $ &1.49	&35.28	&0	\\
			\hline
			
			1300& 2 fat jets with $p_{Tmin}=300$ GeV&  13.66&	1036.8	&1120199.03  \\
			&2 top-tagged jets& 1.55&	137.05&	833.46\\
			&$p_{T}^{l}>580$ GeV and $p_{T}^{sl}>540$ GeV&	0.69 &	30.78	&302.16\\
			& $b$-tagging for both the jets &	0.32 &	15.71&0\\
			&$1280~{\rm{GeV}}<m_{tt}<1400~{\rm{GeV}} $ &0.16 &	3.24	&0	\\
			\hline
		\end{tabular}
	\end{center}
	\caption{\it Cut flow table for two values of the KK Higgs mass.}\label{table1}
\end{table}

\begin{figure}[t!]
	\begin{center}
		\includegraphics[origin=rb,width=3.5in]{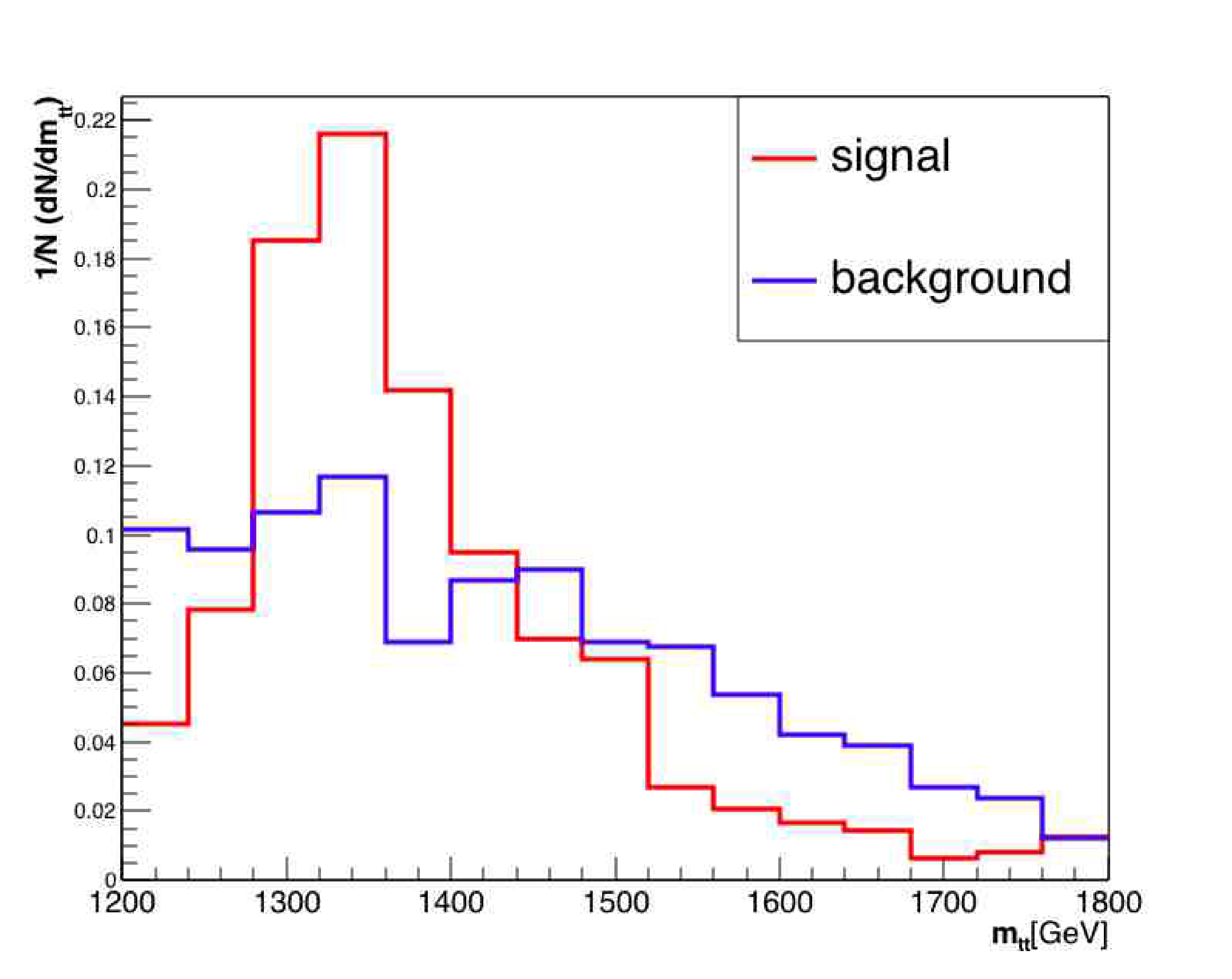}
	\end{center}
	\caption{\it Normalised distributions of the invariant mass ($m_{tt}$) of the leading jet and subleading jet. The red distribution represents the signal and the blue one represents the $t\bar{t}$ background for $M_{h_{1}}=1300~\rm{GeV}$.} \protect\label{mtt}
\end{figure}

Finally we demand that the invariant mass of the $t \bar t$ pair lies within a window close to the $h_{1}$ mass. We find that due to the ISR (Initial State Radiation) the peak of the invariant mass gets smeared towards higher $m_{tt}$ as shown in Figure \ref{mtt}. We find the effect of ISR decreasing as the $h_{1}$ mass increases.

\begin{table}
	\begin{center}
		\begin{tabular}{|c|c|c|} \hline
			$Mh_{1}$& Luminosity in fb$^{-1}$ &Luminosity in fb$^{-1}$\\
			(GeV)&    	for 5$\sigma$ result &	for 3$\sigma $ result\\   \hline 
		
			900	&397&	143\\
			1000&	400&	144\\
			1100&	739&	266\\
			1200&	1477&	531\\
			1300&	3166&	1139\\
			\hline
		\end{tabular}
	\end{center}
	\caption{\it Integrated luminosity in {\rm{fb}}$^{-1}$ for 5$\sigma$ and 3$\sigma$ sensitivities.}\label{table2}
\end{table}

\begin{figure}[!t]
	\begin{center}
		\begin{tabular}{cc}	
			\includegraphics[width=8 cm, height= 5cm]{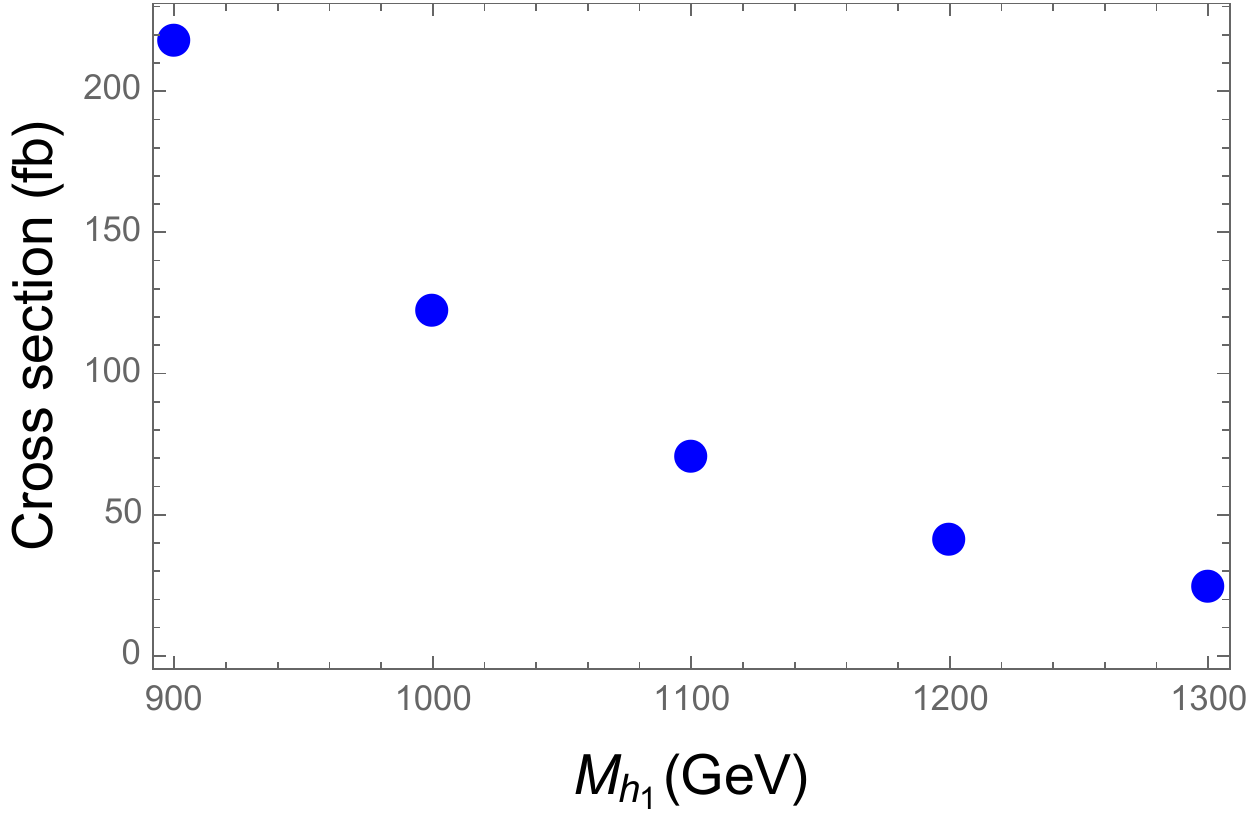}&	
			\includegraphics[width=8 cm, height= 5cm]{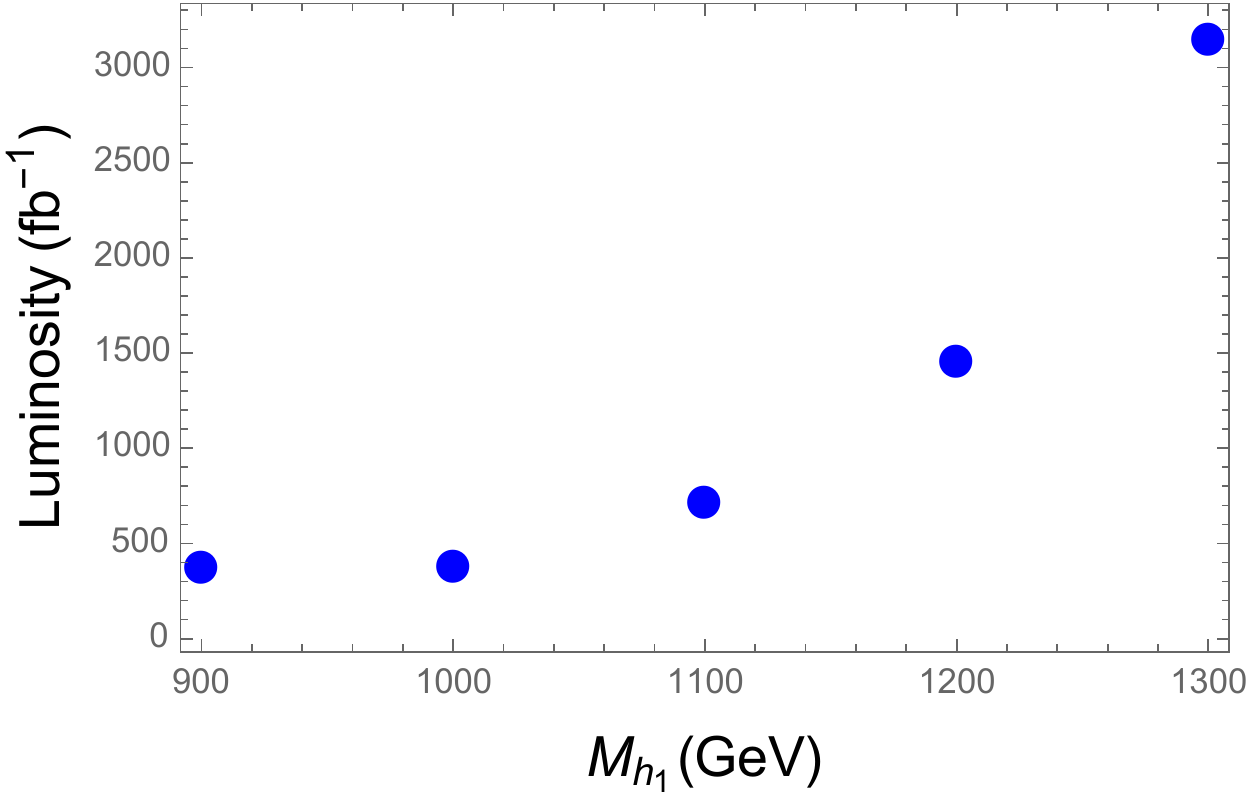}\\
		\end{tabular}
	\end{center}
	\caption{\it The effective cross section for the first mode of the KK Higgs in the deformed RS model is in the left, and the reach of luminosity at the LHC in the right.}\protect\label{reach}
	
\end{figure}

We find from Figure \ref{reach} and Table \ref{table2} that with the choice of the cuts given in Table \ref{table1} we can probe the $h_{1}$ of mass 900 GeV with the luminosity of 397 fb$^{-1}$ but for higher masses such as 1300 GeV we require a luminosity of 3166 fb$^{-1}$.

\section{Conclusion}
		
The first KK mode of the Higgs $(h_{1})$ in the deformed RS model could be as light as 900 GeV  for a choice of deformed model parameters, $a$, $\nu$ and $\Delta$ that can solve the hierarchy problem and be consistent with electroweak precision tests.

For such a value of KK-mass the cross section is sizeable, inspite of the small couplings in this model. However, one has to also contend with a huge SM background, to address which, we propose a new search strategy.

We start by clustering final particles into fat jets and tag them as top-jets using the {\tt{HEPTopTagger}}. This is followed by a $b$-tagging which demands that the $b$ quark be very close to the $b$-like subjet inside the top-jet. This helps us to deal with the QCD background very effectively.
  
Our study shows that using the set of cuts that we propose, $h_{1}$ of mass 900 GeV could be probed at the LHC with a luminosity below 400 fb$^{-1}$. As the $h_{1}$ mass increases the cross section drops further and the required luminosity rises. Higher masses of the $h_{1}$ would need a more refined analysis or the HL-LHC.

\section{Acknowledgements}
N.M and K.S. would like to acknowledge the support of the CNRS LIA (Laboratoire
International Associ\'e) THEP (Theoretical High Energy Physics) and the INFRE-HEPNET
(IndoFrench Network on High Energy Physics) of CEFIPRA/IFCPAR (Indo-French Centre
for the Promotion of Advanced Research). N.M. would like to thank Abhishek Iyer for
discussions and the Department of Theoretical Physics, TIFR for computational resources.
N.M. would also like to gratefully acknowledge hospitality during her visit to IPN Lyon
while this work was in progress. The authors would like to acknowledge the contributions of
Ushoshi Maitra to the initial stages of this work.

\bibliographystyle{JHEP}

\bibliography{biblio}

\end{document}